\documentclass{article}
\usepackage[a4paper, margin=1in]{geometry}
\usepackage{authblk}
\usepackage{geometry} %
\usepackage{braket}
\usepackage[english]{babel}
\usepackage[square,numbers,sort&compress]{natbib}
\usepackage[colorlinks,citecolor=blue]{hyperref}
\expandafter\def\csname ver@subfig.sty\endcsname{}
\usepackage{float}
\usepackage{amsmath}
\usepackage{amssymb}
\usepackage{xcolor}
\usepackage{graphicx}
\usepackage{tikz}
\usetikzlibrary{decorations.pathmorphing}
\usepackage{url}
\usepackage{grffile}
\usepackage{physics}
\usepackage{float}
\usepackage{booktabs}
\usepackage{comment}
\hypersetup{
  colorlinks=true,
  linkcolor=blue,
  filecolor=magenta,
  urlcolor=cyan,
  pdftitle={Overleaf Example},
  pdfpagemode=FullScreen,
}
\usepackage{xspace}
\usepackage{abstract}
\usepackage{titling} %

\newcommand{\be}{\begin{equation}}
\newcommand{\ee}{\end{equation}}
\newcommand{\bea}{\begin{eqnarray}}
\newcommand{\eea}{\end{eqnarray}}
\newcommand{\ben}{\begin{enumerate}}
\newcommand{\een}{\end{enumerate}}
\newcommand{\bp}{\begin{pmatrix}}
\newcommand{\ep}{\end{pmatrix}}
\newcommand{\bb}{\begin{bmatrix}}
\newcommand{\eb}{\end{bmatrix}}
\newcommand{\bit}{\begin{itemize}}
\newcommand{\eit}{\end{itemize}}

\makeatletter
\def\@maketitle{%
  \begin{center}%
  \let \footnote \thanks
    {\LARGE \@title \par}%
    \vskip 0.5em%
    {\normalsize
      \lineskip .5em%
      \begin{tabular}[t]{c}%
        \@author
      \end{tabular}\par}%
    \vskip 0.5em%
    {\normalsize \@date}%
  \end{center}%
  \par
  \vskip 1em}
\makeatother
\begin{document}
\title{\textbf{Quantum Error Correction in Quaternionic Hilbert Spaces}}

\author{Valentine Nyirahafashimana\textsuperscript{1,3*}, Nurisya Mohd Shah\textsuperscript{1,2**}, Umair Abdul Halim\textsuperscript{4}, Mohamed Othman\textsuperscript{1,5},\\  Sharifah Kartini Said Husain\textsuperscript{1,6}
}

\affil{\textsuperscript{1}Institute for Mathematical Research, Universiti Putra
Malaysia,\\ 43400 Serdang, Selangor,
Malaysia}
\affil{\textsuperscript{2}Department of Physics, Faculty of Science, Universiti Putra
Malaysia,\\ 43400 Serdang, Selangor,
Malaysia}

\affil{\textsuperscript{3}Kigali Independent University (ULK), Polytechnic Institute, Kigali Campus,\\ 102 KG 14 Ave Gisozi, Rwanda}

\affil{\textsuperscript{4}Centre of Foundation Studies in Science, Universiti Putra Malaysia,\\ 43400 UPM Serdang, Selangor, Malaysia}
\affil{\textsuperscript{5}Department of Communication Technology and Network, Universiti Putra Malaysia,\\ 43400 UPM Serdang,  Selangor, Malaysia}
\affil{\textsuperscript{6}Department of Mathematics and Statistics, Faculty of Science, Universiti Putra Malaysia,\\ 43400 UPM Serdang, Selangor, Malaysia\\
\textsuperscript{*} Corresponding author: nyirahafashimanav@ulk.ac.rw \\
\textsuperscript{**} Corresponding author: risya@upm.edu.my}

\date{\today}

\maketitle

\begin{center}
  \LARGE{\textbf{Abstract}}
\end{center}

We propose quaternion-based strategies for quantum error correction by extending quantum mechanics into quaternionic Hilbert spaces. Building on the properties of quaternionic quantum states, we define quaternionic analogues of Pauli operators and quantum gates, ensuring inner product preservation and Hilbert space conditions. A simple encoding scheme maps logical qubits into quaternionic systems, introducing natural redundancy and enhanced resilience against noise. We construct a quaternionic extension of the five-qubit code, introducing a framework of 15 syndrome measurements to detect quaternionic errors, including quaternionically rotated error components. Numerical estimates show that the quaternionic five-qubit code achieves a logical error threshold of approximately \(p_{th} \approx 0.015\), demonstrating improved performance compared to the standard complex-valued code. These results suggest a new pathway for quantum error correction in high-noise environments, leveraging the richer structure of quaternionic quantum mechanics to improve fault tolerance.

\section{Introduction}

Quantum error correction (QEC) is essential for realizing fault-tolerant quantum computing, protecting fragile quantum states from errors such as bit flips, phase flips, and amplitude damping ~\cite{steane1998introduction,devitt2013quantum}. Logical qubits are typically formed by entangling multiple physical qubits using codes such as the surface code~\cite{fowler2012surface} being a widely studied example. Recent advances include the use of reconfigurable atom arrays to build logical quantum processors~\cite{bluvstein2024logical}, where neutral atoms serve as physical qubits. This work demonstrated scalable improvements by increasing the surface code distance from $d = 3$ to $d = 7$. Similarly, for superconducting processor achieved below-threshold error correction, showing a logical error suppression factor of $\lambda = 2.14 \pm 0.02$ by increasing the code distance~\cite{acharya2024quantum}. These developments represent key steps toward scalable quantum computation.

Quaternion quantum mechanics (QQM) extends standard quantum mechanics by using quaternions instead of complex numbers. Yang~\cite{yang1950} showed that, in a generalized quantum framework, pure states can be uniquely represented as rays in a vector space. This requires the Hilbert space $\mathcal{H}$ to be defined in one of the three number fields~\cite{danielewski2020foundations}:
\begin{itemize}
    \item $\mathbb{R}$: real numbers
    \item $\mathbb{C}$: complex numbers
    \item $\mathbb{H}$: quaternions
\end{itemize}
Quaternions were introduced as a natural extension of complex numbers into four dimensions. Quaternions, introduced by Hamilton~\cite{hamilton1843new}, extend complex numbers to four dimensions and are defined as\(q = w + xi + yj + zk\), where imaginary units satisfy the relations: $i^2 = j^2 = k^2 = ijk = -1$, to represent quantum states and operators~\cite{voight2021quaternion}. Here \(q\in \mathbb{H}\), \(w,x,y,z \in  \mathbb{R} \) and \(xi + yj + zk\)  is a vector.

Adler~\cite{adler1995quaternionic} developed a complete formulation of quantum mechanics in quaternionic Hilbert spaces, aiming for a possible unification with gravity. Later, Danielewski and Sapa~\cite{danielewski2020foundations} derived QQM from a Cauchy elastic continuum model, proposing it as a physically motivated alternative. Their work introduces quaternionic versions of non-relativistic and relativistic wave equations, but does not address QEC. Although QQM is theoretically well established, its application to QEC remains largely unexplored. Existing research on surface codes and color codes~\cite{fowler2012surface}, and on continuous-variable systems~\cite{aoki2009quantum}, does not consider quaternionic frameworks. In principle, QEC methods such as logical encoding and syndrome measurement can be extended to quaternionic Hilbert spaces, which satisfy the same core postulates as complex ones. However, Vlasov~\cite{vlasov1999error} showed that certain QEC schemes fail in standard complex quantum mechanics but succeed in real and quaternionic cases. Adler also suggested the feasibility of quaternionic quantum field theories~\cite{alder1986quaternionic}. Whether quaternions can offer practical advantages for QEC, such as improved error detection or fault tolerance, remains an open question.

In this work, we investigate the mathematical foundations and applications of quaternionic quantum mechanics (QQM) for quantum error correction. By leveraging the richer algebraic structure of quaternions, we define quaternionic analogues of quantum gates and Pauli operators, and construct new error correction schemes. In particular, we extend the five-qubit code into a quaternionic Hilbert space, introducing 15 syndrome measurements to detect quaternionic errors. We also compare the logical error performance of the standard and quaternionic codes, showing improved resilience against noise and decoherence. 
The paper is structured as follows:
Section~\ref{sec:II} introduces the mathematical foundations of QQM, emphasizing quaternion algebra in error detection.
Section~\ref{sec:III} presents quaternionic Pauli-like operators, error classification, and syndrome mapping.
Section~\ref{sec:IV} compares the logical error rates of the standard and quaternion-extended five-qubit codes.
Section~\ref{sec:V} concludes with a summary and future directions.

\section{Mathematical Framework of Quaternion Quantum Mechanics}\label{sec:II}
In this section, we introduce a new mathematical framework for QQM that ensures both theoretical consistency and practical use in quantum computation. It focuses on five main components: the structure of quaternionic Hilbert spaces, unitary evolution and operators, a basic encoding scheme, and the implementation of quantum gates, specifically the Hadamard and CNOT gates, and building a quaternionic \([[5,1,3]]\) Code with H-Qubits . While based on existing developments, our approach includes new extensions to improve both formal clarity and computational practicality.

\subsection{Quaternionic Hilbert Space}

In standard quantum mechanics, states are vectors in a complex Hilbert space \( \mathcal{H}_\mathbb{C} \). In QQM, we define a Quaternionic Hilbert Space \( \mathcal{H}_\mathbb{H} \) over the quaternion algebra \( q\).
 A state \( \ket{\psi} \in \mathcal{H}_\mathbb{H} \) is expressed as:
\begin{equation}
\ket{\psi} = \sum_n q_n \ket{n}, \quad q_n := w_n + x_ni + y_nj + z_nk,\end{equation}
where  \(w_n, x_n, y_n, z_n \in \mathbb{R}\), \( \{ \ket{n} \} \) is an orthonormal basis, and \( q_n \in \mathbb{H} \).

\noindent

In standard quantum mechanics, where probability amplitudes are complex, the quaternionic structure allows for a more general representation of quantum states. However, the interpretation of probabilities in this framework requires a well-defined inner product which reduces the standard complex form when quaternions are restricted to a complex sub-algebra. The inner product in \( \mathcal{H}_\mathbb{H} \) must be quaternion-valued and satisfy positivity and linearity. Define:
\[
\braket{\phi}{\psi} = \sum_n \overline{\phi_n} q_n,
\]
where the quaternion conjugate is \( \overline{\mathcal{\phi}_n} = w_n - x_ni - y_nj - z_nk, \quad \phi\in \mathcal{H}_\mathbb{H}\) and \(\overline{\mathcal{\phi}_n}\in \mathbb{H} \),  ensuring a real-valued probability interpretation. The norm is defined as:
\begin{equation}
\norm{\psi}^2 = \text{Re}(\braket{\psi}{\psi}) = \sum_n (w_n^2 + x_n^2 + y_n^2 + z_n^2),\label{norm}
\end{equation}
ensuring \( \norm{\psi}^2 \geq 0 \), with equality only if \( \ket{\psi} = 0 \).

To account for the non-commutative nature of quaternions in inner product spaces, right-linearity is imposed:  
\[\braket{\phi}{\psi q } = \braket{\phi}{\psi} \,q , \quad \text{for } q \in \mathbb{H}.\]  
Normalization is then performed using the real part of the inner product, ensuring compatibility with physical interpretations while maintaining the integrity of the quaternionic framework.

\subsection{Unitary Evolution and Operators}

The operators in \( \mathcal{H}_\mathbb{H} \) are quaternion-linear transformations \( A: \mathcal{H}_\mathbb{H} \to \mathcal{H}_\mathbb{H} \) satisfying:
\[
A(\ket{\psi} q) = (A \ket{\psi})q.\]
The adjoint \( A^\dagger \) is defined via:
\[\braket{\phi}{A \psi} = \braket{A^\dagger \phi}{\psi}.\]

Unitary evolution in quaternionic quantum mechanics is governed by a generalized Schrödinger equation:
\[
i\hbar \frac{d}{dt} |\psi\rangle = H |\psi\rangle,
\]
where \( H \) is a Hamiltonian with quaternion value. The challenge in this formulation is ensuring that the evolution remains norm-preserving, which requires a careful definition of unitary operators \( U \) satisfying \( U U^{\dagger} = I \) in the quaternionic Hilbert space. The evolution of time is governed by a unitary operator \( U \), and the Schrödinger equation takes the form:
\[
i \frac{d}{dt} \ket{\psi(t)} = H \ket{\psi(t)},
\]
where \( H = H^\dagger \) is self-adjoint and \( i \) is fixed as the imaginary unit. The solution is:
\[
\ket{\psi(t)} = U(t) \ket{\psi(0)}, \quad U(t) = e^{-i H t},
\]
with the exponential defined via a convergent power series over \( \mathbb{H} \).
We introduce a \emph{quaternion phase alignment} rule: constraining \( U \) to commute with \( i \), implies that \( [U, i] = 0 \), reducing ambiguity from non-commutativity and maintaining compatibility with standard evolution.

\subsection{Simple Encoding Scheme}

A \textit{quaternionic qubit} (q-qubit) is a normalized vector in a two-dimensional quaternionic Hilbert space \( \mathcal{H}_\mathbb{H} \). Its general form is
\begin{equation}
\ket{\psi} = \alpha \ket{0} + \beta \ket{1}, \quad \alpha, \beta \in \mathbb{H}, \quad \text{with } |\alpha|^2 + |\beta|^2 = 1,
\end{equation}
where \( \ket{0} = (1,0) \) and \( \ket{1} = (0,1) \) are the computational basis states. Here, \( \alpha \) and \( \beta \) are quaternions containing four real components each, extending the usual complex qubit structure by including the non-commutative units \( j \) and \( k \). Classical bits are encoded as \( \ket{0} \) and \( \ket{1} \), but the quaternionic structure allows for richer encoding. The \( j \) and \( k \) components can carry additional information, such as global phases, entanglement flags, or hidden logical data. This extra structure increases redundancy and improves robustness against noise. As an example, consider encoding a logical qubit into two physical q-qubits:
\begin{equation}
\ket{\psi_L} = \alpha \ket{00} + \beta i \ket{01} + \gamma j \ket{10} + \delta k \ket{11}, \quad \alpha, \beta, \gamma, \delta \in \mathbb{R},
\end{equation}
where information is spread across the quaternionic units \( i \), \( j \), and \( k \). This distribution introduces natural redundancy: deviations in the \( j \) or \( k \) directions can signal errors and trigger correction procedures. Measurement in QQM projects the state onto the computational basis, similar to standard quantum mechanics, with probabilities
\begin{equation}
P(0) = |\alpha|^2, \quad P(1) = |\beta|^2,
\end{equation}
where \( |\alpha|^2 = \alpha \bar{\alpha} \) and \( |\beta|^2 = \beta \bar{\beta} \) using quaternionic conjugates. Measurement collapses the q-qubit to a classical outcome. Decoding uses quaternionic unitary operations to recover logical information and detect errors by analyzing the imaginary components. This mapping of extra dimensions (\( j \) and \( k \)) to auxiliary data enables compact encoding and more resilient error correction. Such schemes highlight how quaternionic quantum mechanics can model quantum states with richer structures and offer new methods for encoding and protecting information.

\subsection{Quantum Gates and Computation}
In QQM, standard quantum gates can be extended to operate on quaternion-valued state vectors, enriching the expressive capacity of quantum transformations. A quaternionic extension of the Hadamard gate called Hadamard-like gate can be formulated as:
\[
H_\mathbb{H} = \frac{1}{\sqrt{2}}
\begin{bmatrix}
1 & i \\
i & -1
\end{bmatrix},
\]
where \( i \) denotes a quaternionic unit, distinct from the imaginary unit in complex numbers. The quaternionic Hadamard gate introduces non-commutativity through its quaternionic entries, leading to new types of interference and transformation dynamics in quantum states. Alternatively, one can keep the standard matrix form of the Hadamard gate, but apply it to qubits with quaternion-valued amplitudes. Its action on the computational basis remains:
\[
H\ket{0} = \frac{1}{\sqrt{2}} (\ket{0} + \ket{1}), \quad
H\ket{1} = \frac{1}{\sqrt{2}} (\ket{0} - \ket{1}),
\]
where \( \ket{0} \) and \( \ket{1} \) have coefficients in the quaternionic field \( \mathbb{H} \). Yet its application to a general quaternionic qubit state: 
\begin{equation}
\ket{\Psi} = w\ket{0} + (x i + y j + z k)\ket{1}.\label{given state}
\end{equation}
results in
\begin{equation}
\begin{aligned}
\Psi_\mathbb{H}= &H_\mathbb{H} \ket{\Psi} =\frac{1}{\sqrt{2}} (\alpha \ket{0} + \beta \ket{1} ) \quad \text{where}:\\
&\alpha = \frac{1}{\sqrt{2}}(w - x + y k - z j), \quad
\beta = \frac{1}{\sqrt{2}}((w - x)i - y j - z k).
\end{aligned}
\end{equation}
This extension shows that QQM can incorporate richer algebraic structures while keeping quantum operations linear. To test the consistency and computational behavior of quaternionic circuits, we prepare a Bell state using quaternionic gates. This process serves as a minimal benchmark for entanglement generation, gate action on q-qubits, and the preservation of quaternionic amplitudes during evolution.  
We start by applying the quaternionic Hadamard gate \( H_\mathbb{H} \) to the first q-qubit of the initial state \( \ket{00} \). The quaternionic Hadamard generalizes the standard Hadamard by preserving quaternionic coefficients.  
Its action on \( \ket{00} \) produces:
\[
(H_\mathbb{H} \otimes I) \ket{00} = \frac{1}{\sqrt{2}} (\ket{00} + i\ket{10}).
\]
Next, a quaternionic CNOT gate, denoted \( \text{CNOT}_\mathbb{H} \), is applied, which performs a controlled-NOT operation while incorporating quaternionic components in its matrix structure. A possible definition of such a gate is:
\[
\text{CNOT}_\mathbb{H} =
\begin{bmatrix}
1 & 0 & 0 & 0 \\
0 & i & 0 & 0 \\
0 & 0 & 0 & j \\
0 & 0 & k & 0
\end{bmatrix},
\]
where \( i, j, k \) are the fundamental quaternionic units. This matrix applies imaginary quaternionic factors conditionally depending on the control qubit and swaps components in a non-commutative fashion, characteristic of quaternionic extensions. For a given state in~\eqref{given state}: 
\[
\ket{\Psi} = (w\ket{0} + (x i + y j + z k)\ket{1}) \otimes \ket{0}
\]
Applying \( \text{CNOT}_{\mathbb{H}} \) to \( \ket{\Psi} \):
\begin{equation}
\ket{\Psi_{\mathbb{H}}} = \text{CNOT}_{\mathbb{H}} \ket{\Psi}= \alpha\ket{00} + \beta\ket{11}, \quad\text{with} \quad \alpha=w, \quad \beta= y i - x j - z.
\end{equation}

The illustration of the gate’s effect on a general two-q-qubit state, let us consider this state
\[\ket{\Psi'} = ((a + bi) \ket{0} + (c + dj) \ket{1}) \otimes \ket{0}.
\]
Applying a quaternionic CNOT-like gate yields:
\[ \text{CNOT}_\mathbb{H} \ket{\Psi'} =
(a + b i)\ket{00} + ( d i+c k)\ket{11}\]
This result entangles the two q-qubits, distributing quaternionic components across computational basis states.The action of the \(\text{CNOT}_\mathbb{H}\), is defined by:
\[
\text{CNOT}_\mathbb{H}(\ket{\Psi}_c \otimes \ket{\phi}_t) = \ket{\Psi}_c \otimes (\ket{\phi}_t \oplus f(\ket{\Psi}_c)),
\]
where \( f(\ket{\Psi}_c) = 1 \) if the real part of \(\braket{1}{\Psi_c}\) is nonzero, c stands for control qubit subscript and t represents target qubit subscript. This models a conditional flip based on the control qubit's real overlap with \(\ket{1}\).
Combining these operations, we find:
\begin{equation}
\text{CNOT}_{\mathbb{H}} (H_{\mathbb{H}} \otimes I) |00\rangle = \frac{1}{\sqrt{2}} \left( |00\rangle - j |11\rangle \right).
\end{equation}
For the standard Bell state, taking \( j = k = -i \) in \(\text{CNOT}_\mathbb{H}\), this becomes:
\begin{equation}
\frac{1}{\sqrt{2}} \left( |00\rangle + |11\rangle \right),
\end{equation}
embedded in a quaternionic Hilbert space. Quaternionic coefficients \(i, j, k\) are preserved during this evolution, maintaining the non-commutative structure. We then consider the gate set \( \{ H_\mathbb{H}, \text{CNOT}_\mathbb{H}, T \} \), where \( T = \text{diag}(1, e^{i\pi/4}) \) adds a non-Clifford phase needed for universal computation. These gates act linearly on the real subspace while keeping quaternionic coefficients intact, simplifying calculations without full quaternionic matrix operations. This shows that standard quantum protocols, like entanglement generation, extend naturally to the quaternionic setting, offering richer computational structures and new possibilities for error encoding.

\subsection{Quaternionic Construction of the five-qubit Code}

The quaternionic \([[5,1,3]]\) error correction code, built from the codewords \(|00000\rangle\) and \(|11111\rangle\), extends the construction used for three H-qubits~\cite{vlasov1999error}. Under right \(SU(2)\) actions (via quaternionic multiplication) on a single H-qubit, the codeword transformations are:

\begin{equation}
\begin{array}{c|c|c|c}
\text{codeword} & \times i & \times j & \times k \\ 
\hline
|00000\rangle & |10000\rangle & |\dot{0}0000\rangle & |\dot{1}0000\rangle \\
|11111\rangle & -|01111\rangle & -|\dot{1}1111\rangle & |\dot{0}1111\rangle \\
\end{array}
\end{equation}

Here, \(|\dot{0}\rangle\) and \(|\dot{1}\rangle\) represent basis states arising from right multiplication by \(i\), with the mappings \(1 \mapsto |0\rangle\), \(i \mapsto |\dot{0}\rangle\), \(j \mapsto |1\rangle\), and \(k \mapsto |\dot{1}\rangle\).

Expanding each H-qubit into two standard qubits, the transformations in the tensor product space become:

\begin{equation}
\begin{array}{c|c|c|c}
\text{codeword} & \times i & \times j & \times k \\
\hline
|0_00_00_00_00_0\rangle & |1_00_00_00_00_0\rangle & |0_10_00_00_00_0\rangle & |1_10_00_00_00_0\rangle \\
|1_01_01_01_01_0\rangle & -|0_01_01_01_01_0\rangle & -|1_11_01_01_01_0\rangle & |0_11_01_01_01_0\rangle \\
\end{array}
\end{equation}

where the quaternion basis expands as \(1 \mapsto |0_0\rangle\), \(i \mapsto |0_1\rangle\), \(j \mapsto |1_0\rangle\), and \(k \mapsto |1_1\rangle\). The right \(SU(2)\) errors satisfy the generalized quantum error correction conditions under the Euclidean norm. By adding appropriate ancilla H-qubits, these errors can be corrected through an orthogonal procedure, extending the three-H-qubit method to a full \([[5,1,3]]\) quaternionic code.

\section{Application to Error Correction}\label{sec:III}

In standard QEC, stabilizer codes use Pauli operators over complex Hilbert spaces to detect and correct bit-flip ($X$) and phase-flip ($Z$) errors, based on the binary structure of qubit states. QQM extends QEC into quaternionic Hilbert spaces $\mathcal{H}_{\mathbb{H}}$, where quantum states have quaternion-valued amplitudes. This allows for detecting new types of errors, called quaternionic ($\mathcal{Q}$) errors, by incorporating quaternionic rotations and auxiliary measurements. For example, the operator
\begin{equation}
\mathcal{Q}_j = I \otimes (j^\dagger j) \label{qdetector}
\end{equation}
can detect deviations in the $j$- and $k$-components, improving fault tolerance against non-Pauli errors. In this framework, the error model includes not only bit-flip and phase-flip errors but also quaternionic rotational errors, generated by unitary operations like
\[
U_q = e^{i\theta j} = \cos\theta + i \sin\theta j,
\]
which rotate the $j$-component of the state’s amplitudes without changing the basis states. Applying $U_q$ to $\alpha\ket{0}$ produces $(e^{i\theta j}) \alpha \ket{0}$, modifying the internal structure of the quaternionic coefficient. These errors are detected by measuring specific quaternionic components, for example with $\mathcal{Q}_j$, where the expectation value $\langle \mathcal{Q}_j \rangle = c^2$ reveals the strength of the $j$-component.

Physically, quaternionic structures introduce $\text{SU}(2)$-like symmetries that naturally appear in noisy systems, such as optical setups, and relate to quaternionic field theories and topological quantum computing. In this approach, logical qubits called $q$-qubits are encoded within the quaternionic algebra
\begin{equation}
\mathbb{H} = \left\{ q = w + xi + yj + zk \mid w, x, y, z \in \mathbb{R},\ i^2 = j^2 = k^2 = ijk = -1 \right\}. \label{Quaternion}
\end{equation}
This offers more degrees of freedom, providing redundancy and greater resilience in high-noise environments. Building on the foundations of QQM quaternionic Hilbert spaces, unitary evolution, and quaternionic versions of Hadamard and CNOT gates, we propose a quaternionic stabilizer code. As a first step, we construct a 3-$q$-qubit code that encodes a single logical $q$-qubit, inspired by the 3-qubit bit-flip code in standard quantum mechanics. Logical basis states are defined as
\[
\ket{0_L} = \ket{000}, \quad \ket{1_L} = \ket{111},
\]
forming the logical codeword
\begin{equation}
\ket{\psi_L} =  \alpha \ket{000} +  \beta \ket{111}. \label{q-state}
\end{equation}
The code space is a subspace of $\left( \mathcal{H}_{\mathbb{H}} \right)^{\otimes 3}$, its stabilizer group is generated by quaternion-valued operators
\[
S_1 = Z_1 Z_2 I_3, \quad S_2 = I_1 Z_2 Z_3,
\]
which leave both $\ket{0_L}$ and $\ket{1_L}$ invariant, and $Z = \begin{pmatrix} 1 & 0 \\ 0 & -1 \end{pmatrix}$ is the Pauli-Z operator acting on $\{\ket{0}, \ket{1}\}$. Logical operations are defined by
\[
X_L = X_1 X_2 X_3, \quad Z_L = Z_1 Z_2 Z_3,
\]
where $X = \begin{pmatrix} 0 & 1 \\ 1 & 0 \end{pmatrix}$ is the Pauli-X operator. These logical operators act analogously to Pauli operations on the encoded states and are transversal across the three physical qubits. Within the stabilizer formalism, stabilizers are operators that define the code space by commuting with all encoded states and are used for error detection through syndrome measurements. These operators commute with the stabilizers but act non-trivially on the encoded information, representing the logical degrees of freedom. Their action on the logical basis states satisfies
\[
Z_L \ket{0_L} = \ket{0_L}, \quad Z_L \ket{1_L} = -\ket{1_L}, \quad X_L \ket{0_L} = \ket{1_L}, \quad X_L \ket{1_L} = \ket{0_L}.
\]

Syndrome measurement distinguishes errors using the eigenvalues of stabilizer generators. For instance, if an $X$-error occurs on the first $q$-qubit in~\eqref{q-state}, the resulting state is
\[
|\psi \rangle=\alpha \ket{100} +  \beta \ket{011},
\]
and the measured syndromes are $S_1 = -1$, $S_2 = +1$, pinpointing the location of the error. Correction involves applying the appropriate inverse operator: $X$ for bit-flips, and $U_q^\dagger = e^{-i\theta j}$ for quaternionic rotations.
To improve error correction, we extend the five-qubit \([[5,1,3]]\) code~\cite{bennett1996mixed,laflamme1996perfect,yoder2016universal} to quaternionic framework, encoding one logical $q$-qubit into five physical $q$-qubits. The code distance $d = 3$ allows correction of any single-$q$-qubit error and detection of up to two errors. The logical basis states are
\[
\ket{0_L} = \ket{00000}, \quad \ket{1_L} = \ket{11111}.
\]
The stabilizer group is generated by
\[
\begin{aligned}
S_1 &= X_1 X_2 X_3 X_4 I_5, \\
S_2 &= Z_1 Z_2 I_3 I_4 I_5, \\
S_3 &= I_1 Z_2 Z_3 I_4 I_5, \\
S_4 &= I_1 I_2 Z_3 Z_4 Z_5,
\end{aligned}
\]
with logical operators
\[
X_L = X_1 X_2 X_3 X_4 X_5, \quad Z_L = Z_1 Z_2 Z_3 Z_4 Z_5.
\]

Error detection uses a 4-bit syndrome. For instance, a bit-flip error on the first $q$-qubit yields
\begin{equation}
(S_1, S_2, S_3, S_4) = (-1, -1, +1, +1), \quad\text{where}\quad
s_i = \begin{cases}
+1, & \text{if } [E, S_i] = 0 \quad \text{(commute)}, \\
-1, & \text{if } \{E, S_i\} = 0 \quad \text{(anti-commute)}.
\end{cases} \label{syndrome}
\end{equation}
 Here $E$ represents the error operator acting on the code space, $S_i$ is i-th stabilizer generator of the code and $s_i$ is the syndrome bit corresponding to stabilizer $S_i$. Quaternionic errors, which modify internal amplitudes without flipping basis states, produce a trivial syndrome $(+1, +1, +1, +1)$, and detecting such errors requires additional measurements of $j$ and $k$ components. Corrections are applied using $X_i$, $Z_i$, or $U_q^\dagger$, depending on the identified error. This quaternionic extension of the five-qubit perfect code reinterprets Pauli operators, stabilizers using quaternionic structures and explores how syndrome measurements reflect quaternionic error type.

\subsection{Quaternionic Representation of Errors}

In quaternionic QEC, standard Pauli operators \( X, Y, Z \) mapped to quaternion-valued operations, within quaternions set defined in~\eqref{Quaternion}. We define the following Quaternionic Pauli-like Operators, on single \emph{q}-qubit in Table~\ref{tab:quaternionic-pauli}:

\begin{table}[!ht]
\centering
\caption{Quaternionic Extensions of Pauli Operators}
\begin{tabular}{ccl}
\hline
\textbf{Operator} & \textbf{Quaternionic Form} & \textbf{Interpretation} \\
\hline
\( X \) & \( \mathcal{Q}_X = iX \) & Bit-flip with quaternionic phase \\
\( Y \) & \( \mathcal{Q}_Y = jY \) & Bit and phase-flip with quaternionic phase \\
\( Z \) & \( \mathcal{Q}_Z = kZ \) & Phase-flip with quaternionic twist \\
\( I \) & \(\mathcal{Q}_I = I \) & Identity operator \\
\hline
\end{tabular}
\label{tab:quaternionic-pauli}
\end{table}

These quaternionic Pauli-like operators act on the Hilbert space by inducing both linear transformations on qubit states and rotational transformations in the quaternionic phase space. Unlike standard operators, they introduce a non-commutative phase structure, enriching quantum error representation within a four-dimensional (4D) geometric framework.
\subsection{Syndrome and Error Mapping (Quaternionic)}
 
Each single-qubit error is mapped to its corresponding syndrome and extended to a quaternionic form, as illustrated in Table~\ref{tab:quaternionic-syndromes}. Specifically, each error operator \( E = X, Y, Z \) is associated with its quaternionic version \( \hat{E} = qE \), where \( q = i, j, k \) respectively. These quaternionic phases introduce non-commuting dynamics into the error structure. While the syndrome patterns themselves remain unchanged in magnitude, they are now linked to these generalized quaternionic errors. In this quaternionic extension of the five-qubit code, the overall structure of syndrome measurement remains consistent with the standard stabilizer formalism. However, the errors now carry additional quaternionic structure, such as \( i, j, k \) phases, which enrich the error model. This added structure can be useful for representing non-classical correlations, particularly in the context of exotic decoherence channels or non-standard quantum theories where quaternionic formulations may offer greater physical insight.

\begin{table}[!ht]
\centering
\caption{Quaternionic Errors and Syndrome Measurements for the Five-Qubit Code in~\eqref{syndrome}}
\begin{tabular}{cccl}
\hline
\textbf{Error} & \textbf{Syndrome} $(S_1, S_2, S_3, S_4)$ & \textbf{Quaternionic Error} & \textbf{Error Description for each q-qubit} \\
\hline
$X_1$ & $(-1, -1, +1, +1)$ & $iX_1$ & Bit-flip with quaternionic phase on q-qubit 1 \\
$Y_1$ & $(-1, -1, +1, +1)$ & $jY_1$ & Bit and phase-flip with quaternionic phase \\
$Z_1$ & $(+1, -1, +1, +1)$ & $kZ_1$ & Phase-flip with quaternionic phase \\
$X_2$ & $(-1, -1, -1, +1)$ & $iX_2$ & Bit-flip on q-qubit 2 \\
$Y_2$ & $(-1, -1, -1, +1)$ & $jY_2$ & Y-error on q-qubit 2 \\
$Z_2$ & $(+1, -1, -1, +1)$ & $kZ_2$ & Z-error on q-qubit 2 \\
$X_3$ & $(-1, +1, -1, -1)$ & $iX_3$ & Bit-flip on q-qubit 3 \\
$Y_3$ & $(-1, +1, -1, -1)$ & $jY_3$ & Y-error on q-qubit 3 \\
$Z_3$ & $(+1, +1, -1, -1)$ & $kZ_3$ & Z-error on q-qubit 3 \\
$X_4$ & $(-1, +1, +1, -1)$ & $iX_4$ & Bit-flip on q-qubit 4 \\
$Y_4$ & $(-1, +1, +1, -1)$ & $jY_4$ & Y-error on q-qubit 4 \\
$Z_4$ & $(+1, +1, +1, -1)$ & $kZ_4$ & Z-error on q-qubit 4 \\
$X_5$ & $(+1, +1, +1, -1)$ & $iX_5$ & Bit-flip on q-qubit 5 \\
$Y_5$ & $(+1, +1, +1, -1)$ & $jY_5$ & Y-error on q-qubit 5 \\
$Z_5$ & $(+1, +1, +1, -1)$ & $kZ_5$ & Z-error on q-qubit 5 \\
\hline
\end{tabular}
\label{tab:quaternionic-syndromes}
\end{table}
These quaternionic phases enrich the error structure but do not alter the syndrome patterns themselves.\\

\emph{\bf Remark}:
In the quaternionic extension of the five-qubit code, errors such as $iX,jY,kZ$ include additional quaternionic phases. However, since the quaternionic units $i,j,k$ commute with Pauli matrices under scalar multiplication, the syndrome associated with a quaternionic error $qP$ is identical to that of the corresponding standard Pauli error $P$. Therefore, syndrome measurement proceeds in the same manner as in the standard stabilizer formalism, with quaternionic modifications reflected in error correction rather than syndrome detection.\\

This framework naturally generalizes to arbitrary \([[n,k,d]]\) codes within QQM. For such codes, \( n \) denotes the number of physical \( q \)-qubits, \( k \) the number of logical \( q \)-qubits, and \( d \) the code distance, enabling correction of \( t = \lfloor (d-1)/2 \rfloor \) errors. Notable examples include quaternionic adaptations of the Steane \([[7,1,3]]\) and Shor \([[9,1,3]]\) codes. These are extended to support detection and correction of quaternionic errors as include $Q$-error by explicitly tracking their \( j \)- and \( k \)-components. By leveraging the four degrees of freedom inherent in quaternions, such codes achieve efficient error correction with fewer physical qubits compared to their complex-valued counterparts, thereby offering new possibilities in the design of fault-tolerant quantum systems.

\section{Error Rates in Quaternionic Quantum Error Correction}\label{sec:IV}

In standard QEC, the \emph{physical error rate} \(p\) represents the probability of an error occurring on a physical qubit, typically modeled by Pauli operators \(X\), \(Y\), and \(Z\). The \emph{logical error rate} \(p_L\) is the probability that such errors remain uncorrected on the encoded logical qubit. For a code of distance \(d\), logical errors scale approximately as \(p_L \sim p^d\). Surface codes, for example, typically tolerate physical error rates up to a threshold of about \(1\%\)~\cite{fowler2009high}, beyond which reliable error suppression fails.

Quaternionic QEC extends this framework using quaternionic Pauli-like operators (Table~\ref{tab:quaternionic-pauli}), which introduce additional error types such as four-dimensional rotations and non-commutative transformations. These richer dynamics like directional interference and \(4D\) amplitude shifts require adapted decoding strategies and influence both the error threshold \(p_{\text{th}}\) and the scaling behavior of \(p_L\).

QEC aims to preserve logical qubits by encoding them across many entangled physical qubits. When physical errors are below a critical threshold, the logical error rate can be suppressed exponentially with increasing code distance. This behavior is approximated by:
\begin{equation}
p_L \approx \left( \frac{p}{p_{\text{th}}} \right)^{\left\lfloor \frac{d+1}{2} \right\rfloor},
\end{equation}
as established for surface codes~\cite{dennis2002topological, kitaev2003fault, fowler2012surface}. Here \(p\) and \(p_L\) are the physical and logical error rates, \(d\) is the code distance, and \(p_{\text{th}}\) is the threshold error rate, which depend on parameters such as angular resolution \(\theta\) and measurement precision. When \(p \ll p_{\text{th}}\), the logical error rate decreases exponentially with \(d\). The suppression factor,
\[
\lambda = \frac{p_L}{p_{L(d+2)}} \approx \frac{p_{\text{th}}}{p},
\]
quantifies the reduction in logical error rate, and  reflects the improvement in logical fidelity when the code distance increases by two.

\begin{figure}[!ht]
    \centering
    \includegraphics[width=0.5\linewidth]{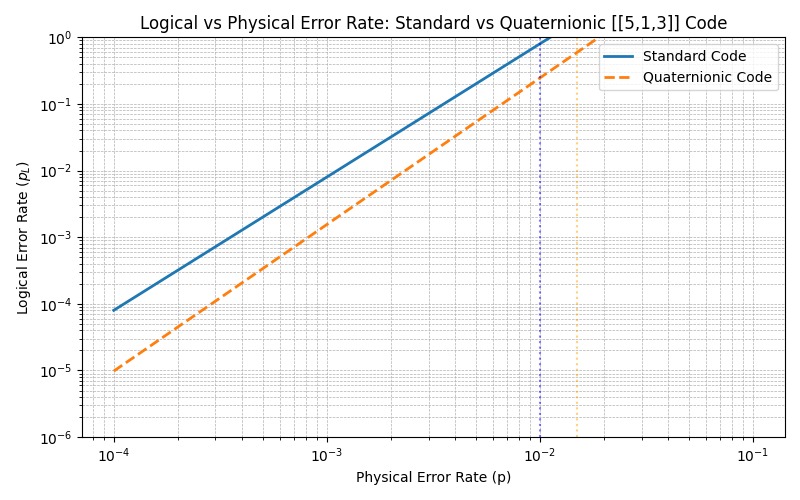}
    \caption{Comparison of the logical error rate versus physical error rate for the standard [[5,1,3]] code and its quaternionic extension.}
    \label{fig:logical}
\end{figure}

As shown in Figure~\ref{fig:logical}, the standard \([[5,1,3]]\) code achieves \(p_L \sim p^2\), reflecting its ability to correct single-qubit errors, and has a threshold near \(p_{\text{th}} \approx 0.01\). Its quaternionic extension supports a broader range of error types, resulting in increased performance: the logical error rate scales as \(p_L \sim p^{2.2}\), with an improved threshold of about \(p_{\text{th}} \approx 0.015\). This improvement stems from the higher-dimensional error distinguishability provided by quaternionic structures. While more powerful codes such as concatenated codes or GKP-encoded hybrids are needed to fully correct all quaternionic error types, the intrinsic structure of quaternionic codes provides advantages in sensitivity and error resilience. These results highlight the potential of quaternionic QEC in advancing fault-tolerant quantum computing.

    
\section{Conclusions}\label{sec:V}

This work develops a quaternion-based formulation of quantum mechanics by extending standard quantum theory to quaternionic Hilbert spaces. It defines quaternionic Pauli-like operators and introduces quantum-like gates (e.g., \(H_\mathbb{H}\), CNOT\(_\mathbb{H}\)). We examine how syndrome measurements in the quaternionic extension of the five-qubit code correspond to quaternionic error components specifically the \(i\), \(j\), and \(k\) directions requiring auxiliary measurements to detect these errors.
The five-qubit quaternionic code achieves a threshold of approximately \(p_{\text{th}} \approx 0.015\), benefiting from improved error distinguishability in higher dimensions. This value is between the thresholds reported in prior studies. For example,~\cite{wootton2012high} reports a threshold of 0.0185 for depolarizing noise, nearing the upper bound of 0.189 for surface codes when noise correlations are considered~\cite{bombin2012strong, ohzeki2012error}. In contrast, Takagi~\cite{takagi2017error} found the five-qubit code exhibits a lower threshold due to its round-robin structure and error accumulation, similar to the Bacon-Shor code.
The proposed framework opens new avenues for QEC by utilizing quaternionic structures to improve noise resilience. While it departs from conventional quantum mechanics, it provides a potential pathway toward more robust and fault-tolerant quantum architectures. Future work will extend this analysis to quaternionic codes \([[n,k,d]]\), with a focus on their syndrome-based error correction properties. The quaternionic \([[5,1,3]]\) code will serve as a baseline case, with generalizations aimed at evaluating feasibility and performance in practical QEC scenarios.

\section*{Acknowledgment}~~ This work is supported by the Air Force Office of Scientific Research under Award No. FA2386-22-1-4062. The authors gratefully acknowledge this support.

\section*{Author Contributions}~~ V.N.: methodology, formal analysis, investigation, writing original draft.
N.M.S.: conceptualization, methodology, investigation, writing - review \& editing, validation, project administration, funding acquisition. U.A.H: co-supervision, visualization, writing - review \& editing.  M.O.: writing - review \& editing. S.K.S.H.: review \& editing.


\bibliographystyle{unsrt}
\bibliography{references}

\begin{thebibliography}{10}

\bibitem{steane1998introduction}
AM~Steane.
\newblock Introduction to quantum error correction.
\newblock {\em Philosophical Transactions of the Royal Society of London.
  Series A: Mathematical, Physical and Engineering Sciences},
  356(1743):1739--1758, 1998.

\bibitem{devitt2013quantum}
Simon~J Devitt, William~J Munro, and Kae Nemoto.
\newblock Quantum error correction for beginners.
\newblock {\em Reports on Progress in Physics}, 76(7):076001, 2013.

\bibitem{fowler2012surface}
Austin~G Fowler, Matteo Mariantoni, John~M Martinis, and Andrew~N Cleland.
\newblock Surface codes: Towards practical large-scale quantum computation.
\newblock {\em Physical Review A—Atomic, Molecular, and Optical Physics},
  86(3):032324, 2012.

\bibitem{bluvstein2024logical}
Dolev Bluvstein, Simon~J Evered, Alexandra~A Geim, Sophie~H Li, Hengyun Zhou,
  Tom Manovitz, Sepehr Ebadi, Madelyn Cain, Marcin Kalinowski, Dominik
  Hangleiter, et~al.
\newblock Logical quantum processor based on reconfigurable atom arrays.
\newblock {\em Nature}, 626(7997):58--65, 2024.

\bibitem{acharya2024quantum}
Rajeev Acharya, Dmitry~A Abanin, Laleh Aghababaie-Beni, Igor Aleiner, Trond~I
  Andersen, Markus Ansmann, Frank Arute, Kunal Arya, Abraham Asfaw, Nikita
  Astrakhantsev, et~al.
\newblock Quantum error correction below the surface code threshold.
\newblock {\em Nature}, 2024.

\bibitem{yang1950}
C.~N. Yang.
\newblock On quantized space-time.
\newblock {\em Physical Review}, 72(9):874--874, 1950.

\bibitem{danielewski2020foundations}
Marek Danielewski and Lucjan Sapa.
\newblock Foundations of the quaternion quantum mechanics.
\newblock {\em Entropy}, 22(12):1424, 2020.

\bibitem{hamilton1843new}
William~Rowan Hamilton.
\newblock On a new species of imaginary quantities, connected with the theory
  of quaternions.
\newblock {\em Proceedings of the Royal Irish Academy (1836-1869)}, 2:424--434,
  1840.

\bibitem{voight2021quaternion}
John Voight.
\newblock {\em Quaternion algebras}.
\newblock Springer Nature, 2021.

\bibitem{adler1995quaternionic}
Stephen~L Adler.
\newblock {\em Quaternionic quantum mechanics and quantum fields}, volume~88.
\newblock Oxford University Press, 1995.

\bibitem{aoki2009quantum}
Takao Aoki, Go~Takahashi, Tadashi Kajiya, Jun-ichi Yoshikawa, Samuel~L
  Braunstein, Peter Van~Loock, and Akira Furusawa.
\newblock Quantum error correction beyond qubits.
\newblock {\em Nature Physics}, 5(8):541--546, 2009.

\bibitem{vlasov1999error}
Alexander~Yu Vlasov.
\newblock Error correction with euclidean qubits.
\newblock {\em arXiv preprint quant-ph/9911074}, 1999.

\bibitem{alder1986quaternionic}
Stephen~L Alder.
\newblock Quaternionic quantum field theory.
\newblock {\em Communications in Mathematical Physics}, 104:611--656, 1986.

\bibitem{bennett1996mixed}
Charles~H Bennett, David~P DiVincenzo, John~A Smolin, and William~K Wootters.
\newblock Mixed-state entanglement and quantum error correction.
\newblock {\em Physical Review A}, 54(5):3824, 1996.

\bibitem{laflamme1996perfect}
Raymond Laflamme, Cesar Miquel, Juan~Pablo Paz, and Wojciech~Hubert Zurek.
\newblock Perfect quantum error correcting code.
\newblock {\em Physical Review Letters}, 77(1):198, 1996.

\bibitem{yoder2016universal}
Theodore~J Yoder, Ryuji Takagi, and Isaac~L Chuang.
\newblock Universal fault-tolerant gates on concatenated stabilizer codes.
\newblock {\em Physical Review X}, 6(3):031039, 2016.

\bibitem{fowler2009high}
Austin~G Fowler, Ashley~M Stephens, and Peter Groszkowski.
\newblock High-threshold universal quantum computation on the surface code.
\newblock {\em Physical Review A—Atomic, Molecular, and Optical Physics},
  80(5):052312, 2009.

\bibitem{dennis2002topological}
Eric Dennis, Alexei Kitaev, Andrew Landahl, and John Preskill.
\newblock Topological quantum memory.
\newblock {\em Journal of Mathematical Physics}, 43(9):4452--4505, 2002.

\bibitem{kitaev2003fault}
A~Yu Kitaev.
\newblock Fault-tolerant quantum computation by anyons.
\newblock {\em Annals of physics}, 303(1):2--30, 2003.

\bibitem{wootton2012high}
James~R Wootton and Daniel Loss.
\newblock High threshold error correction for the surface code.
\newblock {\em Physical review letters}, 109(16):160503, 2012.

\bibitem{bombin2012strong}
H{\'e}ctor Bombin, Ruben~S Andrist, Masayuki Ohzeki, Helmut~G Katzgraber, and
  Miguel~A Martin-Delgado.
\newblock Strong resilience of topological codes to depolarization.
\newblock {\em Physical Review X}, 2(2):021004, 2012.

\bibitem{ohzeki2012error}
Masayuki Ohzeki.
\newblock Error threshold estimates for surface code with loss of qubits.
\newblock {\em Physical Review A—Atomic, Molecular, and Optical Physics},
  85(6):060301, 2012.

\bibitem{takagi2017error}
Ryuji Takagi, Theodore~J Yoder, and Isaac~L Chuang.
\newblock Error rates and resource overheads of encoded three-qubit gates.
\newblock {\em Physical Review A}, 96(4):042302, 2017.

\end{thebibliography}

\end{document}